\pdfoutput=1
\documentclass[reprint, pre, nofootinbib, aps]{revtex4-2}
\usepackage{graphicx} 
\usepackage{float}
\usepackage{amsmath}
\usepackage{amssymb}
\usepackage{soul}
\usepackage[colorlinks=true]{hyperref}
\usepackage{bm}
\usepackage{url}
\usepackage{color}

\begin{document}


\title{Predicting Excitation Energies in Warm Dense Matter}

\author{T. Q. Thelen}
\affiliation{Los Alamos National Laboratory, P.O. Box 1663, Los Alamos, NM 87545, U.S.A.}

\author{D. A. Rehn}
\affiliation{Los Alamos National Laboratory, P.O. Box 1663, Los Alamos, NM 87545, U.S.A.}

\author{C. J. Fontes}
\affiliation{Los Alamos National Laboratory, P.O. Box 1663, Los Alamos, NM 87545, U.S.A.}

\author{C. E. Starrett}
\email{starrett@lanl.gov}
\affiliation{Los Alamos National Laboratory, P.O. Box 1663, Los Alamos, NM 87545, U.S.A.}

\date{\today}

\begin{abstract}

In a dense plasma environment, the energy levels of an ion shift relative to the isolated ion values. This shift is reflected in the optical spectrum of the plasma and can be measured in, for example, emission experiments. In this work, we use a recently developed method of modeling electronic states in warm dense matter to predict these level energies. In this model, excited state energies are calculated directly by enforcing constrained one-electron occupation factors, thus allowing the calculation of specific transition and ionization energies. This model includes plasma effects self-consistently, so the effect of continuum lowering is included in an \emph{ab-initio} sense. We use the model to calculate the K-edge and K-alpha energies of solid density magnesium, aluminum, and silicon over a range of temperatures, finding close agreement with experimental results. We also calculate the ionization potential depression (IPD) to compare to widely used models, and investigate the effects of temperature on the lowering of the continuum.
\end{abstract}

\maketitle

\section{Introduction}
\begin{figure}[!ht]\centering
    \includegraphics[width=0.4\textwidth]{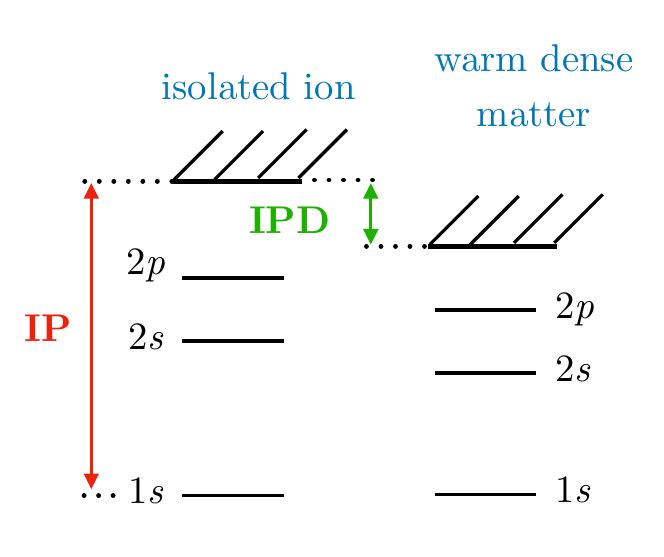}
    \caption{Schematic showing the effects of continuum lowering on the energy levels of an ion in warm dense matter, compared with the energy levels of an isolated ion. The ionization potential (IP) is shown in red, and the ionization potential depression (IPD) is shown in green.}
    \label{fig:ipd-diagram}
\end{figure}

In a hot dense plasma environment, the screening cloud of free electrons causes the energy levels of an ion to shift. A schematic of this energy shift is shown in Fig.~\ref{fig:ipd-diagram}. The ability to model such conditions is crucial in the field of astrophysics, as these conditions can be found in stellar and planetary interiors \cite{astroopacity, planetinteriors, zeng2020}. Also, the opacity of such plasmas is the subject of current experimental efforts \cite{perry2017replicating, nagayama2019systematic}. The lowering of the continuum, or ionization potential depression (IPD), in dense plasmas affects atomic binding energies, as well as the cross-sections of atomic processes, such as collisional ionization and excitation. It can also limit the number of allowed bound states, which shifts the charge state distribution towards higher ionization, affecting the thermodynamic properties of the system, such as the equation of state and opacity. 

Recent experiments carried out at the Linac Coherent Light Source (LCLS) \cite{OC2012, oc16}, the Orion laser \cite{hoarty}, and the National Ignition Facility (NIF) \cite{Fletcher} have measured the IPD in warm dense matter. In addition to the IPD, some of these experiments also measured K-edge and K-alpha energies.

A number of sophisticated models have been used recently to predict the IPD. These include a two-step Hartree-Fock method \cite{sks}, classical molecular dynamic simulations \cite{Calisti}, Monte-Carlo simulations \cite{stransky}, simulations based on finite-temperature DFT \cite{vinko2014, hu2017}, atomic-solid-plasma models \cite{rosmej2018}, and average-atom DFT methods \cite{JCP}. Some of these models have also been used to calculate K-edge and K-alpha energies. 

Two well-known and widely used IPD models are the Stewart-Pyatt (SP) \cite{SP} and Ecker-Kroll (EK) \cite{EK} models, which are popular due to being rapid to evaluate and simple to implement. However, these models need to be tested and verified through comparisons with more sophisticated models and experiments. The validity of these models is important, as many atomic physics codes implement them in their calculations. For example, the Stewart-Pyatt \cite{SP} model is used in the Los Alamos suite of atomic physics codes~\cite{fontes2015} and in the FLYCHK~\cite{flychk} code.


In this work, we use our recently developed excited states model (ESM) \cite{starrett24excited} to predict the IPD, and compare with the SP and EK models. Our model has plasma screening built in, so it includes the IPD without an external, \emph{ad hoc} model. In addition to the IPD, we use our model to calculate the K-edge and K-alpha energies to compare to experiments \cite{OC2012, oc16}. The model is found to make accurate predictions for the experiments on magnesium, aluminum, and silicon. We also discuss limitations of our model, including the local density approximation (LDA) for exchange and correlation, and the lack of multiple-scattering.

We first present a summary of our model in Section~\ref{sec:ESM}. In Section~\ref{sec:kedge}, we use this model to calculate K-edge and K-alpha energies, and compare our findings with experimental results. In Section~\ref{sec:IPD}, we calculate the IPD and compare with the SP and EK models.

\section{Excited States Model \label{sec:ESM}}
We first review the ESM, originally presented in Ref.~\cite{starrett24excited}, and use the atomic model presented in that work. The strategy of this model is to calculate the energies of individual excited states using effective single-electron expressions. The individual excited states are defined by a chosen set of single-particle level occupations. For example, we may wish to calculate the energy of a $1s^2 2s^2 2p^6+FD$ configuration in magnesium. This notation means that we fix the occupation of the $1s$ shell to be 2, the $2s$ shell to be 2, the $2p$ shell to be 6, and the remaining 2 electrons have Fermi-Dirac (FD) occupation factors.  

The definition of a particular orbital is such that it extends into the continuum. This means, for example, that if an excited state has 2 electrons in the $3s$ orbital, but the $3s$ orbital is not bound, or is partially bound, the definition smoothly assigns part of the continuum to be the $3s$ orbital. Such a smooth definition of the orbitals is necessary to avoid discontinuities in physical quantities such as pressure and temperature changes. See Ref. \cite{starrett24excited} for details. 

The model is based on minimizing the free energy that explicitly includes the excited states. In Hartree atomic units, the free energy is written
\begin{equation}
\begin{split}
    F = & \sum_x W_x [E_x - T S_x] +T \sum_x W_x \log W_x \,,
\end{split}
\end{equation}
where the sum is over all non-degenerate (in energy) electronic excited states, $x$, of the system; $W_x$ is the probability of the excited state; $E_x$ is the internal energy; $S_x$ is the entropy of the electrons in the excited state; and $-W_x \log W_x$ is the entropy associated with energy-degenerate excited states.

The energy is given by the effective single-particle expression
\begin{equation}
    E_x = E_x^{el} +  E_x^{xc} +  E_x^{(0)} \,,
\end{equation}
in which $E_x^{el}$ is the electrostatic energy, $ E_x^{xc}$ is the exchange and correlation internal energy, and $ E_x^{(0)}$ is the kinetic energy of the electrons. The electrostatic energy, $E_x^{el}$, is given by
\begin{equation}
    E_x^{el} = \frac{1}{2} \int_V n_{x}(r) \left[ V_x^{el}(r) - \frac{Z}{r} \right] \; d^3r \,,
\end{equation}
where $V$ is the volume of the ion sphere (which is equal for all excited states and determined by the atomic mass and mass density), $Z$ is the nuclear charge, $n_x(r)$ is the electron density associated with excited state $x$, and the electrostatic potential is
\begin{equation}
    V_x^{el}({r}) = -\frac{Z}{r} + \int_{V} \frac{n_{x}(r')}{|\bm{r} - \bm{r'}|} \; d^3r' \,.
\end{equation}
The exchange and correlation energy is given by 
\begin{equation}
    E_x^{xc,LDA} = \int_{V}  \epsilon^{xc}\left[n_{x}(r)\right]  \; d^3r \,.
\end{equation}
The kinetic energy of electrons in the configuration is 
\begin{equation} 
\begin{split}
    E_x^{(0)} = \sum_{i\in B} n_{x,\epsilon_i,l_i} N_{x,\epsilon_i,l_i} \epsilon_{i} \; +& 
    \int_{0}^{\infty} n_{x,\epsilon,l}\, \epsilon \,N_{x,\epsilon,l} \; d\epsilon \\
    -& \int_{V} \left( V_x^{eff}(r) - \gamma\right) n_{x}(r) \; d^3r \,.
    \label{Equation 9}
\end{split}
\end{equation}
The effective single-particle potential $V_x^{eff}(r)$ is
\begin{equation}
    V_x^{eff}({r}) = V_x^{el}({r}) + V_x^{xc}({r}) \,,
\end{equation}
in which $V_x^{xc}(r)$, the exchange and correlation potential, is given by
\begin{equation}
    V_x^{xc}({r}) = \frac{\delta F^{xc}}{\delta n_{x}({r})} \,,
\end{equation} 
where $F^{xc}$ is the chosen exchange and correlation free energy.
The normalization factor, $N_{x,\epsilon,l}$, in Eq.~\ref{Equation 9} is given by the integral
\begin{equation}
    N_{x,\epsilon,l} = \int_0^R y_{x,\epsilon,l}(r)^2 \; dr
    \label{Equation 11}
\end{equation}
and the constant $\gamma$ sets the zero of energy
\begin{equation}
\begin{split}
    \gamma =& \sum_x W_x V_x^{xc}(R)\label{gamma} \,,
\end{split}
\end{equation}
where $R$ is the ion sphere radius.

In a practical sense, the model starts with an initial guess at the effective one-particle potential, $V_x^{eff}(r)$, for a given configuration $x$, and uses that to solve the radial Schr\"odinger equation for the eigenfunctions, $y_{x,\epsilon,l}(r)$,
\begin{equation}
    \frac{d^2y_{x,\epsilon,l}(r)}{dr^2} + 2 \left( \epsilon_{x,l} - V_x^{eff}(r) - \frac{l(l+1)}{2r^2}  +\gamma \right)y_{x,\epsilon,l}(r) = 0 \,,
\end{equation}
in which $\epsilon_{x,l}$ is the eigenenergy of the eigenstate in configuration $x$ and orbital angular momentum quantum number $l$. The model then constructs the electron density
\begin{equation}
\begin{split}
    n_{x}(r) =& \sum_{i\in B}  2(2l_i+1)\frac{n_{x,\epsilon_i, l_i}}{4\pi r^2} y_{x,\epsilon_i,l_i}(r)^2  \\
    &+ \int_0^{\infty}  \sum_{l=0}^{\infty} 2(2l+1) \frac{n_{x,\epsilon,l}}{4\pi r^2} y_{x,\epsilon,l}(r)^2 
 \; d\epsilon
\end{split}
\end{equation}
with occupations $n_{x,\epsilon,l}$ of configuration $x$, and the first term is a sum over all bound states $B$.  The occupation factors, or occupation numbers, $n_{x,\epsilon,l}$ are set at input, and ensure that the ion sphere is charge neutral, i.e.
\begin{equation}
    Z - \int_{V} n_{x}({r}) \; d^3r =0 \,,
\end{equation} 
where $Z$ is the nuclear charge. Using the determined $n_{x}(r)$, a new effective potential is formed, and the process is repeated until self-consistency is achieved.

A key choice in this model is the list of excited states, defined by a set of occupation factors $\{n_{x,\epsilon,l}\}$.  Following Ref.~\cite{starrett24excited}, the list of excited states includes integer permutations of the occupations of one-electron orbitals up to a chosen $n_{max}$. For example, for $n_{max}=2$, we consider integer permutations of the occupations of the $1s$, $2s$, and $2p$ orbitals. Any remaining electrons that are not in these shells are given Fermi-Dirac occupations. The physical reasoning for choosing $n_{max}$ is based on the lifetime of the orbital. If an experimental time scale is much longer than an excitation lifetime, then the measurement will be averaged over such excited states. For example, excitations in free electron states are very short-lived due to rapid collisional decay, whereas excitations of core electrons are relatively long-lived.

\section{K-edge and K-alpha \label{sec:kedge}}

\begin{figure}[H]
    \centering
    \includegraphics[width=0.47\textwidth]{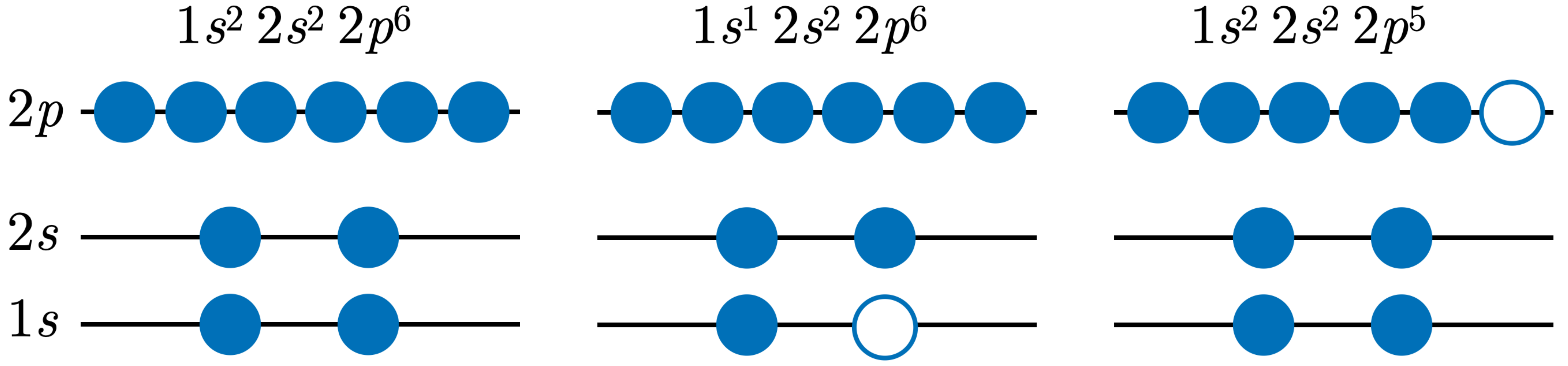}
    \caption{Schematic diagram showing all of the configurations needed to calculate the K-edge and K-alpha for magnesium with 10 bound electrons (charge state 2). On the left is the ground state.  In the middle, an electron has been ionized from the K-shell. The difference between the energies of these two configurations gives the K-edge energy. On the right, an electron from the $2p$ has  radiatively decayed to the K-shell. The difference in energies between this ion and the middle ion gives the K-alpha energy.}
    \label{fig:kconfigs}
\end{figure}

The K-edge refers to the minimum amount of energy needed to ionize an electron from the $1s$ orbital, or the K-shell. Since the energy levels of an ion in warm dense matter are shifted from that of an isolated ion, the K-edge energy will also be different. This is true for transition energies as well. The K-alpha transition occurs when an electron from the $n=2$ shell radiatively decays to the K-shell, after an electron has been ionized from the K-shell. This transition emits a photon, and the energy of that photon is the K-alpha energy (see Fig.~\ref{fig:kconfigs}). 

A series of experiments \cite{oc16, OC2012} provided experimental measurements of the K-edge and K-alpha for solid density magnesium, aluminum, and silicon. In these experiments, conducted with the Linac Coherent Light Source (LCLS), an x-ray free-electron laser (XFEL) was used to excite and probe a target with a range of laser energies. If the energy of the laser was great enough, an electron from the K-shell would be ionized. Then, an electron from the L-shell ($n=2$ orbital) would radiatively decay, filling the hole in the K-shell and emitting a photon. The energy of the emitted photons was measured in the experiment, as well as the intensity of the emission. During this process, the target heated up due to the thermalization of the ionized electrons. Temperature was not measured, but, due to the short timescales (hundreds of femtoseconds), the plasma mass density could be assumed to be unchanged (i.e., remain at solid density).

\begin{figure}
    \centering
    \includegraphics[scale=0.17]{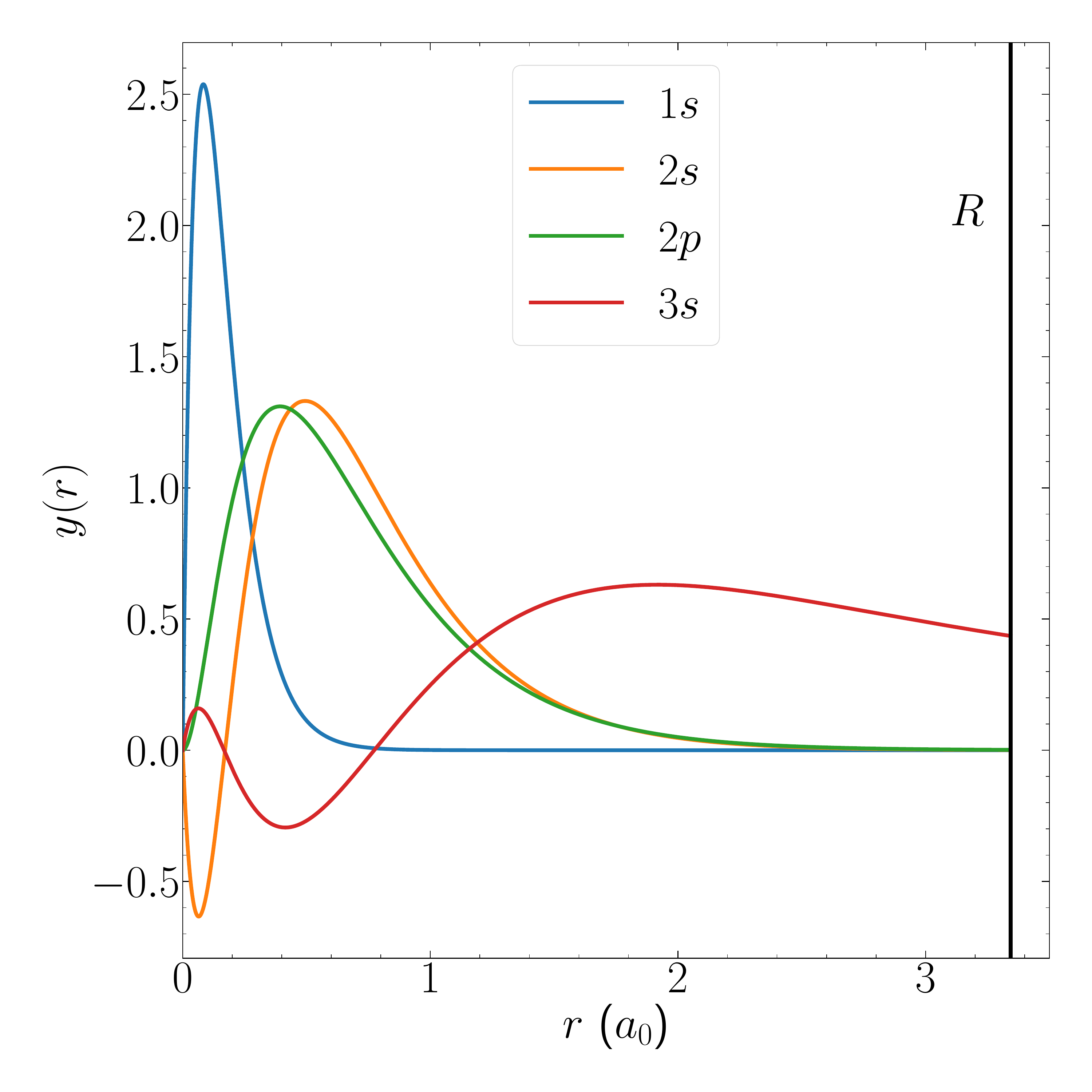}
    \caption{Radial wavefunctions $y(r)$ of a magnesium configuration inside the ion sphere with radius $R=3.34\; a_0$. Due to it not being well localized inside the ion sphere, the $3s$ orbital is deemed to be hybridized, and is assumed to be occupied with Fermi-Dirac statistics. The other eigenstates are well localized and we permute of all possible integer occupations of them to create a configuration list.}
    \label{fig:wfs}
\end{figure}

To model these experiments with the present model, we set $n_{max}=2$ for all three materials. This choice was based on the degree of localization of the $n=3$ wavefunctions. As seen in Fig.~\ref{fig:wfs}, the $3s$ wavefunction is not well contained within the Wigner-Seitz radius $R$; it is hybridized. This indicates that electrons in this state will be itinerant, like free electrons. Excitations will therefore have short lifetimes, and it is reasonable to use $n_{max}=2$. 

\begin{figure}
    \centering
    \includegraphics[scale=0.17]{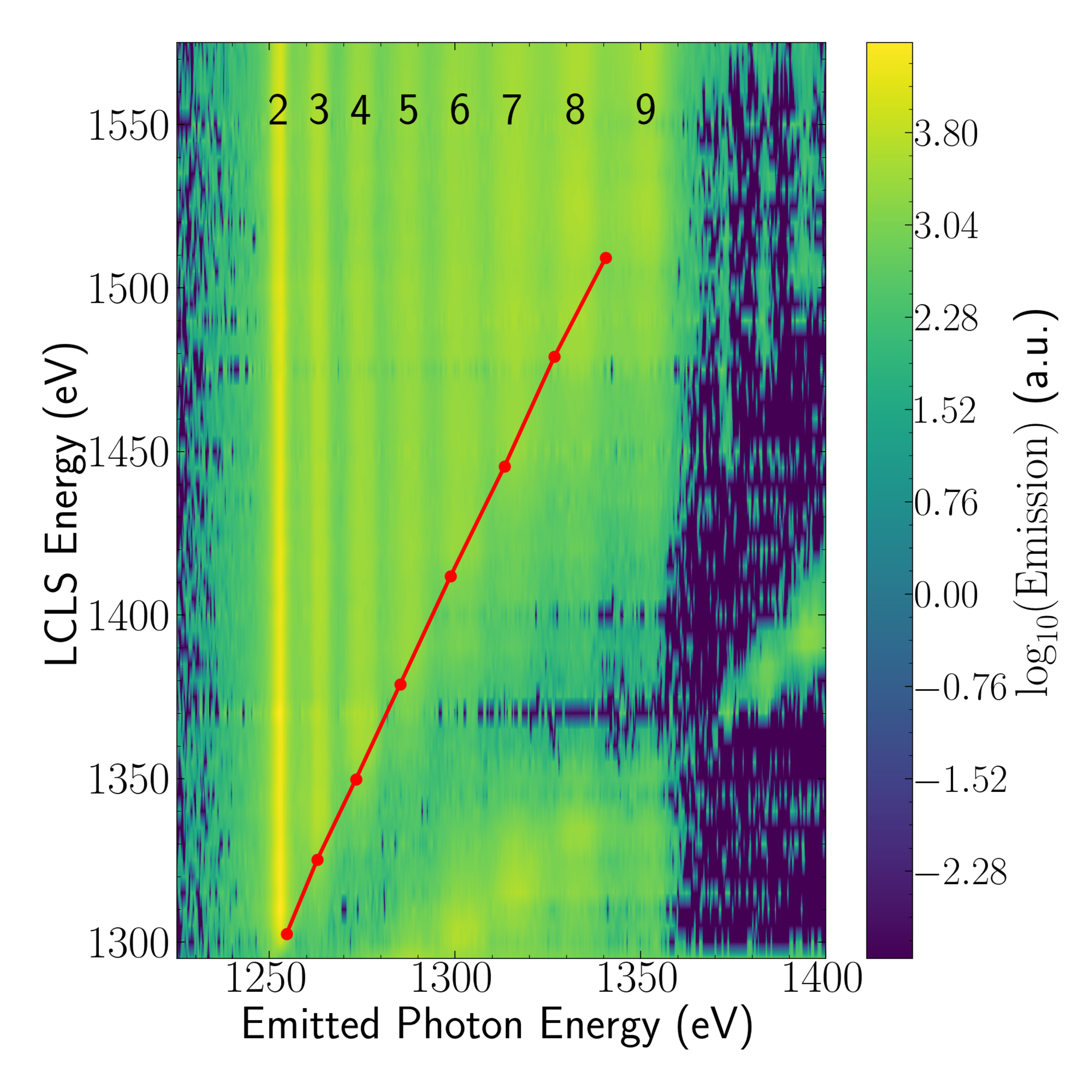}
    \caption{K-edge and K-alpha results calculated using our model for magnesium at solid density plotted atop experimental results. The experimental results \cite{oc16} show the measured intensity of emission on a logarithmic scale in arbitrary units. The charge states corresponding to each K-alpha transition are labeled at the top.}
    \label{fig:contour}
\end{figure}

Fig.~\ref{fig:contour} shows the experimental data from Ref.~\cite{oc16} for solid density magnesium. The different colors reflect the intensity of measured photon emission in arbitrary units, the energy of which is plotted on the horizontal axis. The appearance of the bright vertical lines shows the energy of the K-alpha transitions for different charge states. These vertical lines only begin above a certain LCLS energy. This is due to the fact that the K-alpha transition will only occur once the LCLS energy is enough to ionize an electron from the K-shell. The LCLS energy at which these vertical lines begin for each charge state is the K-edge energy. 

We calculated the K-edge and K-alpha energies for magnesium at solid density using our model. The energies of specific configurations were calculated to find these ionization and transition energies. For example, the ground state of a magnesium ion with ten bound electrons is $1s^2 \; 2s^2 \; 2p^6 + FD$. After an electron is ionized from the K-shell, the new configuration is $1s^1 \; 2s^2 \; 2p^6 + FD$. After an electron from the L-shell fills the hole in the K-shell, the configuration becomes $1s^2 \; 2s^2 \; 2p^5 + FD$. The K-edge energy is then given by
\begin{equation}
    K_{edge} = E_{1s^1 2s^2 2p^6+FD} - E_{1s^2 2s^2 2p^6+FD} - E_{therm}
    \label{Equation 12}
\end{equation}
and the K-alpha energy is given by
\begin{equation}
    K_{\alpha} = E_{1s^2 2s^2 2p^5+FD} - E_{1s^1 2s^2 2p^6+FD} \,,
\end{equation}
in which $E$ is the energy of a given configuration. A visual schematic of the configurations used in this example of magnesium with ten bound electrons is shown in Fig.~\ref{fig:kconfigs}. For the contents of this paper, the label of ``charge state" refers to the charge state of the initial configuration before an electron from the K-shell is ionized. For example, in Fig.~\ref{fig:kconfigs}, the initial configuration (shown on the left) has an ionic charge of 2, while the other two configurations have an ionic charge of 3. The K-edge energy is the difference in energy between the initial configuration with an ionic charge of 2 and the subsequent configuration with an ionic charge of 3, so we label it the K-edge for charge state 2. While the K-alpha energy calculation is the difference in energy between the two configurations with ionic charges of 3, it is labeled as charge state 2, as that is the charge of the initial configuration.

The thermalization energy, $E_{therm}$, is the energy required to thermalize a free electron. This energy is included in the energy of a final-state configuration, as the ionized electron is modeled with the Fermi-Dirac distribution. Since we only want to know the minimum energy required to remove an electron from the K-shell and place it in the continuum, we subtract $E_{therm}$ in Eq.~\ref{Equation 12}. The thermalization energy was calculated assuming a free electron gas model for the ionized electron. 

\begin{figure}
    \centering
    \includegraphics[scale=0.17]{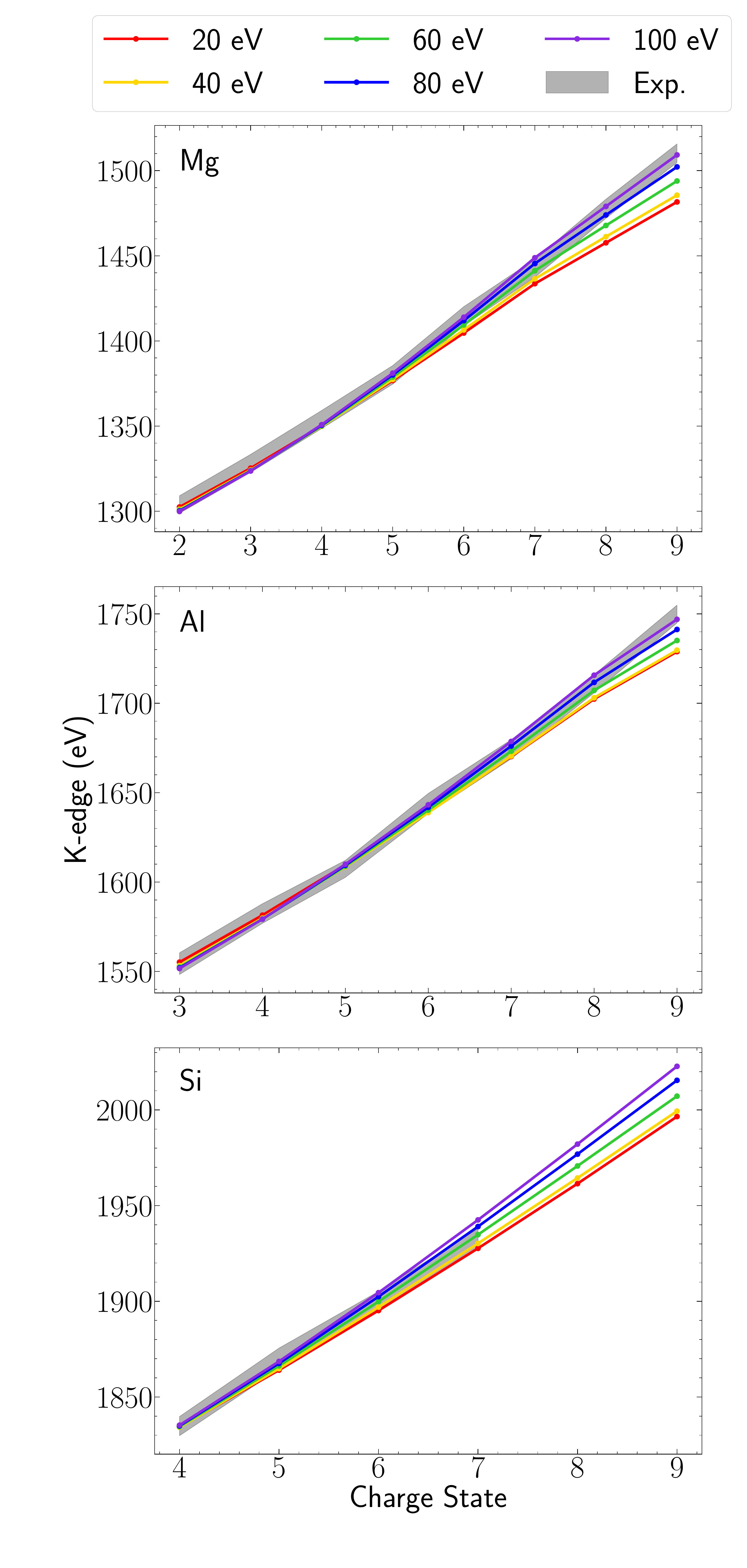}
    \caption{K-edges calculated with our model for temperatures of 20, 40, 60, 80, and 100~eV for solid density magnesium (top), aluminum (middle), and silicon (bottom), compared with experimental results (shaded) \cite{oc16}. Experimental K-edges for silicon were only reported up to charge state 7.}
    \label{fig:kedge-all-elements}
\end{figure}

Fig.~\ref{fig:contour} shows calculated K-edge and K-alpha results compared with the experimental emission results. For this overall comparison, we have chosen the temperature that best fits the K-edge measurements; we address the temperature dependence next. While we see reasonable agreement from the contour plot alone, it is difficult to discern exactly where the K-edge ionizations and K-alpha transitions occur. Ciricosta et. al. \cite{oc16} extracted the K-edge values from the data, to which we compare our results directly. They also extracted the K-edge values for aluminum and silicon, and we compare the predictions of our model for these cases as well. The comparisons of the K-edge energies for all three elements are shown in Fig.~\ref{fig:kedge-all-elements}. 

In Fig.~\ref{fig:kedge-all-elements}, we show K-edge results from our model for several temperatures. Importantly, the temperature of the plasma was not directly measured in the experiments~\cite{oc16}. The predicted K-edge energies become sensitive to temperature at higher charge states. For all elements, the calculated K-edge values for lower charge states are very close to each other, despite being calculated over a large range of temperatures. However, at higher charge states, the calculated K-edge values vary more over the same range of temperatures. This is due to a larger fraction of ionized electrons that thermalize and, therefore, become more sensitive to temperature. Agreement between the model and the experiments is excellent, for all three materials. Even though temperature is not measured, the relative insensitivity of the K-edge values means that the experiment provides a useful constraint on the model.

\begin{figure}
    \centering
    \includegraphics[scale=0.17]{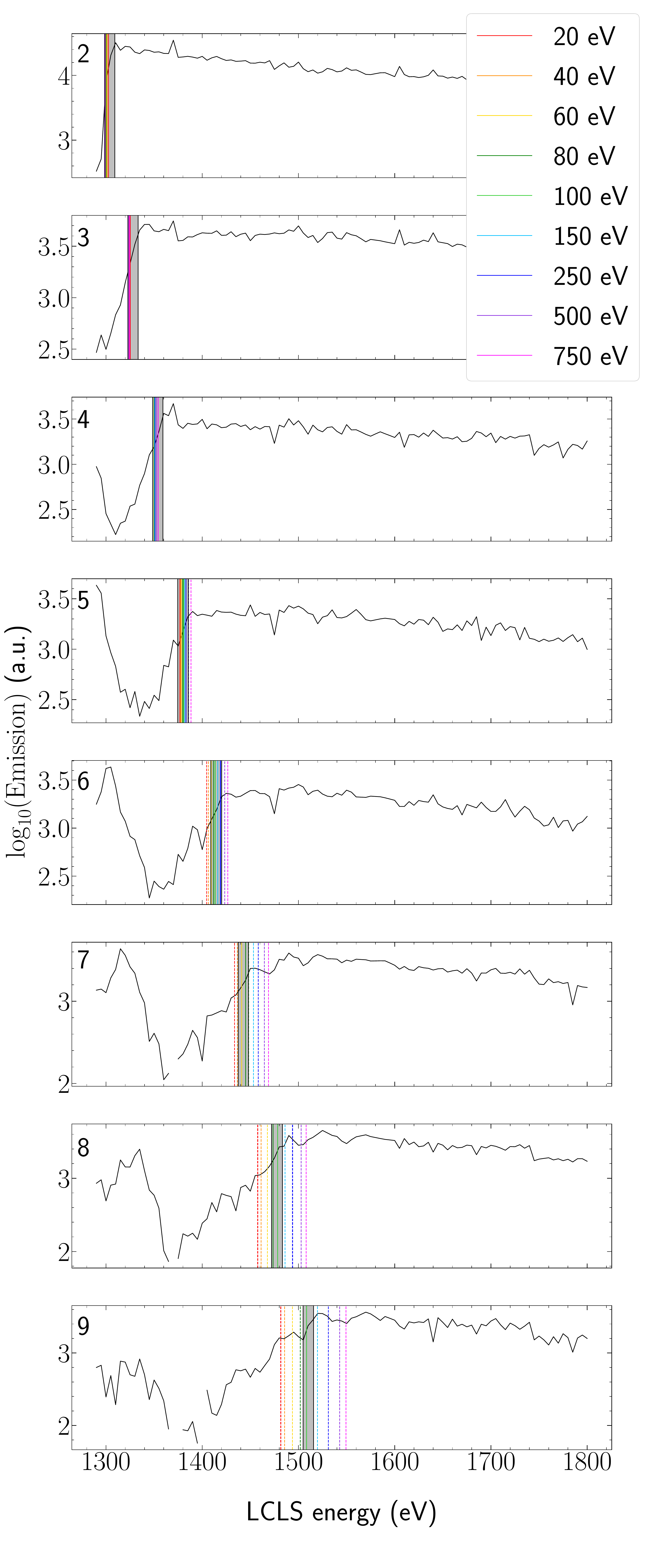}
    \caption[]{Lineouts of experimental data \cite{oc16} taken at peak emitted photon energy for each charge state (labeled in the top left corners), compared with the K-edges calculated by our model at various temperatures (represented by colored vertical lines---solid for falling within experimental range, dashed for falling outside), and the experimental edge (shaded region). The gaps in the lineout data are due to negative values of the emission.}
    \label{fig:kedge-lineout}
\end{figure}

For comparison, Fig.~\ref{fig:kedge-lineout} shows lineouts of the measured experimental emission for magnesium, showing our calculated K-edges over a range of temperatures for each charge state. For lower charge states, there exist very sharp edges in emission, indicating clear values for the  K-edge locations. We see that as the charge state increases, the slope that indicates the edge gets shallower. This is clearly apparent at higher charge states, as we see that the ``edges" for these higher charge states span a much larger range of LCLS energies. The shallower slope of the K-edge reflects two things: first, for higher charge states, bound states relocalize due to the higher charge of the ion, leading to more states near the bound-free threshold; second, the Fermi edge becomes blurred due to the higher temperatures \cite{OC2012}. This behavior makes it more challenging to identify a single value for the K-edge, as the K-edge itself is blurred out.

Our aluminum K-edge results, shown in Fig.~\ref{fig:kedge-all-elements}, also agree with those of the finite-temperature DFT methods presented in Ref. \cite{vinko2014}, as well as the two-step Hartree-Fock method in Ref. \cite{sks}. Our aluminum K-edge is mostly consistent with the K-edge presented in Ref. \cite{iglesias}, however the K-edge calculated with our model is lower than Ref.~\cite{iglesias} for higher charge states. As seen in our results, the higher charge states are more sensitive to temperature variation, so this disagreement could be sensitive to the temperature treatment in the model. In all of our K-edge calculations for all three elements shown in Fig.~\ref{fig:kedge-all-elements}, the K-edge increases with charge state, which is consistent with the aforementioned experiments and models, as well as with the atomic-solid-plasma model presented in Ref.~\cite{rosmej2018}, in which they show the K-edge of a carbon plasma increasing with charge state.

\begin{figure}
    \centering
    \includegraphics[scale=0.17]{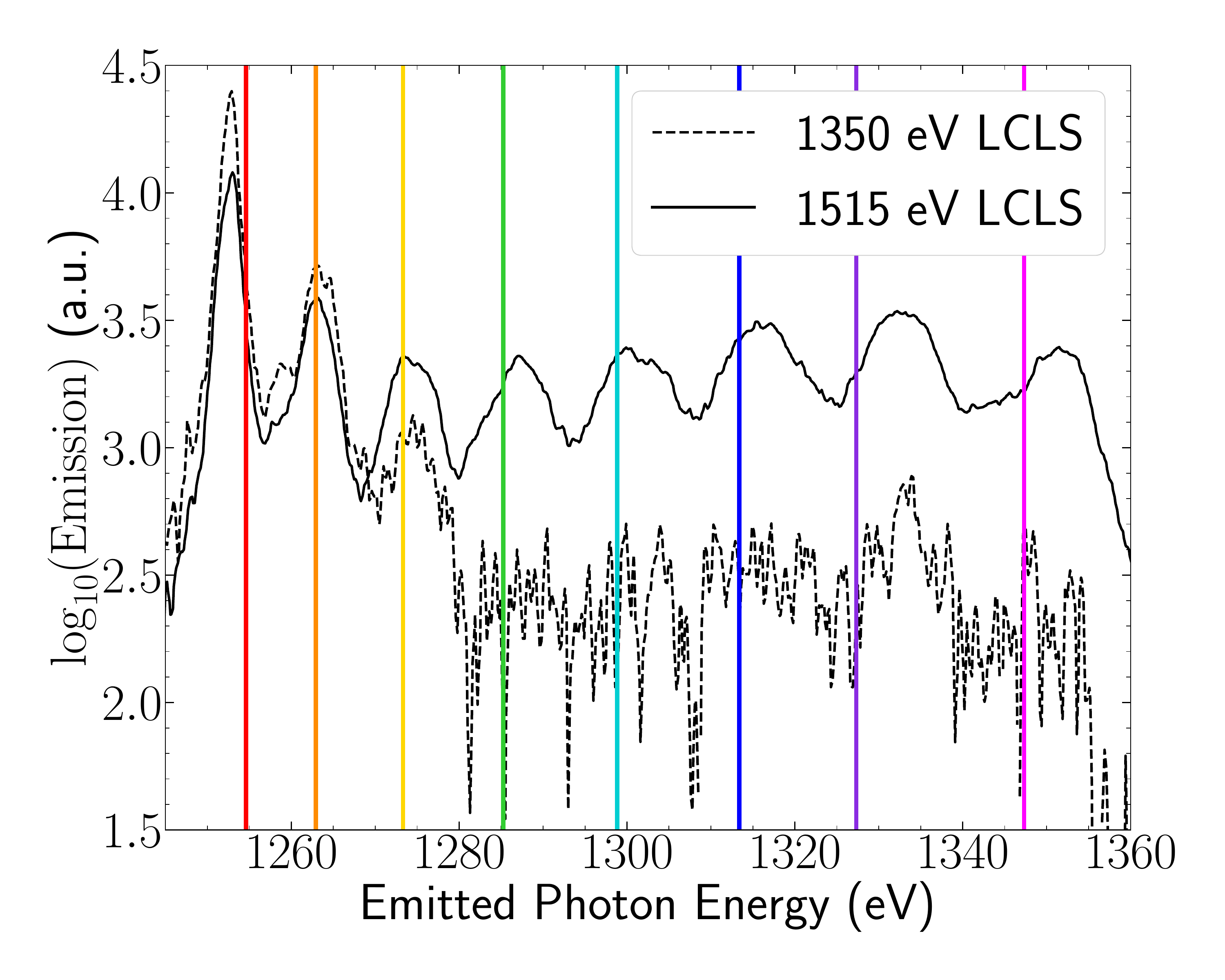}
    \caption{Lineouts of the emission (in arbitrary units) measured in the experiment for magnesium~\cite{oc16} taken at LCLS energies of 1350~eV and 1515~eV. Each peak corresponds to the energy of the photon emitted from a K-alpha transition, with the transition at charge state 2 having the lowest energy and charge state 9 having the greatest energy. The K-alpha  energies calculated with our model are indicated by the colored, solid vertical lines. The lineout at an LCLS energy of 1350 eV only shows peaks for the first three charge states because the LCLS energy is not high enough to ionize an electron from the K-shell in higher charge states. Thus, there is no subsequent K-alpha transition. The 1350 eV lineout shows that the location of the peaks do not significantly differ for different LCLS energies.}
    \label{fig:kalpha-lineout}
\end{figure}

In Fig.~\ref{fig:kalpha-lineout}, we look at lineouts of the experimental data for magnesium, highlighting the K-alpha peaks. We see that the K-alpha values calculated with our model correspond with the peaks in emission for most of the charge states. The K-alpha values plotted in Fig.~\ref{fig:kalpha-lineout} are calculated at a specific temperature for each charge state. This temperature corresponds to the best fit for the K-edge results. Agreement between our model and the data is generally good. However, our K-alpha values for the higher charge states, notably charge states 8 and 9, are $\sim$10~eV from the peaks of the experimental data. Similar findings were also obtained for aluminum by Son et. al.~\cite{sks}, who observed that the K-alpha energies of the higher charge states  were lower than experimental values.

A possible reason for the discrepancy at higher charge states could be because our model does not include multiple scattering \cite{starrett2020multiple}. Multiple scattering will more strongly affect the higher charge states since it has the strongest effect on valence states, which are hybridized, and are more abundant in higher charge states.  Valence states are more abundant due to increased state localisation caused by less screening of the nucleus by core state.  More tightly bound states are less affected by multiple scattering due to their localized nature~\cite{starrett2020multiple}. Our model also does not include configuration interaction, but rather uses LDA for exchange and correlation, which has some error associated with it, and this error could be dependent on the charge state \cite{wilson1995deficiency}.

\section{Ionization Potential Depression \label{sec:IPD}}

The  IPD is the change in ionization energy in a plasma at non-zero temperature and density, relative to the isolated ion case (see Fig.~\ref{fig:ipd-diagram}). We can use our model to calculate the ionization potential of a given ion for arbitrary temperatures and densities. For example, the ionization potential (IP) of a magnesium ion with 10 bound electrons (charge state 2) at temperature $T$ and density $\rho$, assuming $n_{max}=2$,  is given by 
\begin{multline}
    IP(T,\rho) = E(T,\rho)_{1s^2 2s^2 2p^6+FD}  \\
     - (E(T,\rho)_{1s^2 2s^2 2p^5+FD} - E(T,\rho)_{therm}) \,.
\end{multline}

Similarly, the same ionization potential can be calculated for an isolated ion (denoted by $II$). The ionization potential of an isolated magnesium ion with 10 bound electrons is given by
\begin{multline}
    IP(II) = E(II)_{1s^2 2s^2 2p^6+FD} \\
    - (E(II)_{1s^2 2s^2 2p^5+FD} - E(II)_{therm}) \,.
\label{Equation 15}
\end{multline}

The IPD is obtained by taking the difference of the ionization potential for an ion in a hot dense plasma, at some finite temperature and density, with the corresponding ionization potential for an isolated ion. Therefore, the IPD at temperature $T$ and density $\rho$ is given by
\begin{equation}
  IPD = IP(T,\rho) - IP(II) \,.
\end{equation}

\begin{figure}
    \centering
    \includegraphics[scale=0.17]{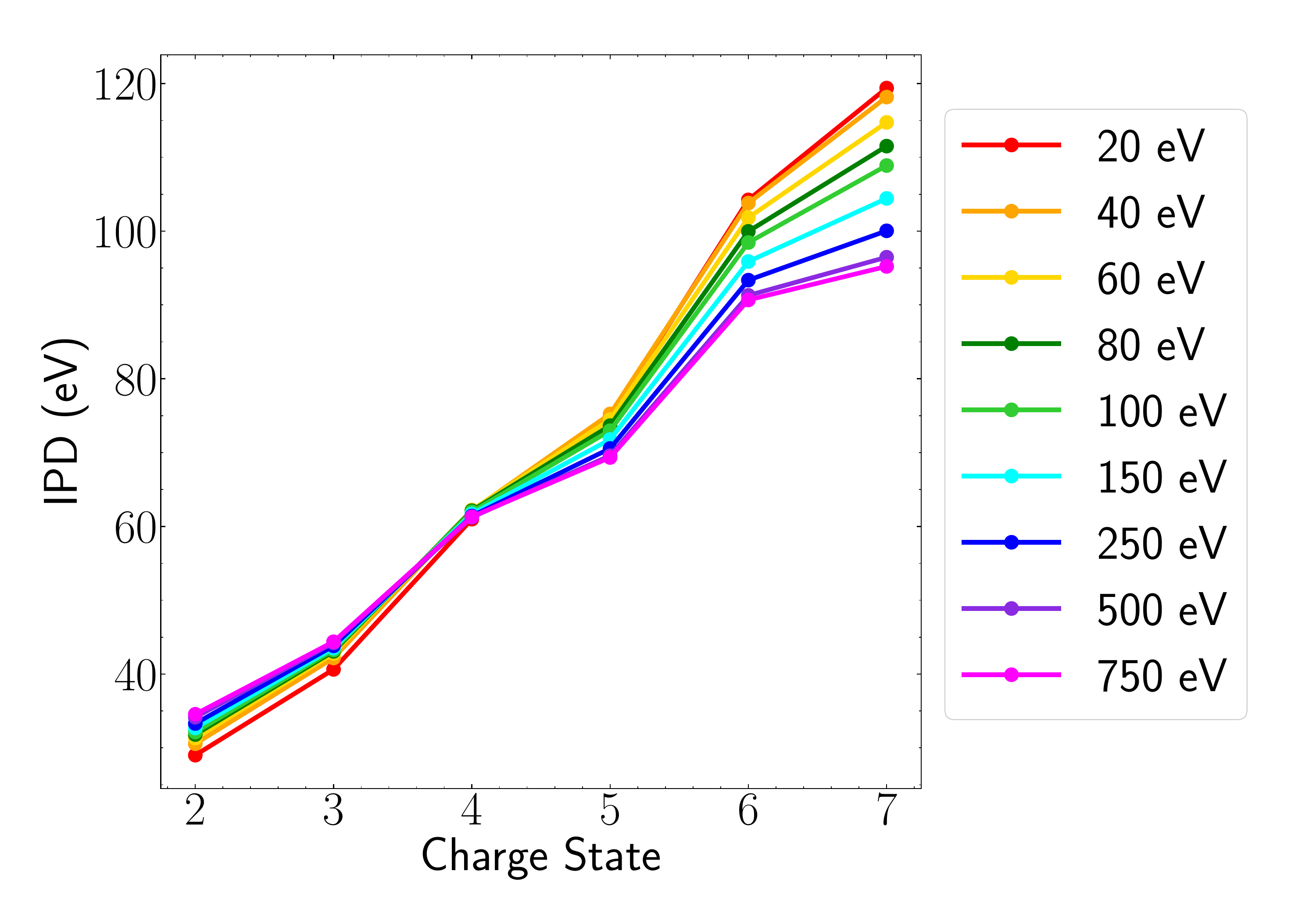}
    \caption{IPDs calculated with our model for temperatures ranging from 20~eV to 750~eV for magnesium at solid density. At charge state 2, the IPD is lowest at a temperature of 20 eV and highest at 750~eV. At a charge state of 7, the IPD is lowest at a temperature of 750~eV and highest at 20~eV.}
    \label{fig:ipd-allT}
\end{figure}

In Fig.~\ref{fig:ipd-allT}, we show the predicted IPD for solid density magnesium plasmas over a range of temperatures. We see that the IPD increases with the charge of the ion, and that it also depends on the temperature of the plasma. Since core states are more weakly affected by plasma conditions than  valence and continuum states, it is to be expected that higher charge states would have a larger IPD than the lower charge states. 

The SP~\cite{SP} and EK~\cite{EK} models are commonly used analytic models of the IPD. The SP IPD is given in Hartree atomic units by
\begin{equation}
    SP = \frac{T}{2(z^*+1)} \left[ \left[ \frac{6z(\pi n_e)^{1/2} (z^*+1)^{3/2}}{T^{3/2}} +1 \right]^{2/3} -1 \right] \,,
    \label{eq:SP}
\end{equation}
where $z$ is the charge state $+1$, $z^*$ is the average ionization of the plasma, $n_e$ is the electron density given by $z^*/V$, and $T$ is the temperature in atomic units. Also in Hartree atomic units, the EK IPD is
\begin{equation}
    EK = \frac{z}{4\pi \epsilon_0} \left[ \frac{4\pi (n_e + n_i)}{3} \right]^{1/3} 
    \label{Equation 18} \,,
\end{equation}
where $z$ and $n_e$ are defined in the same way as the SP model, and $n_i$ is the ionic density, given by $1/V$.

The SP model uses a finite temperature potential for the average electrostatic potential near the nuclei. The free electrons in the plasma are described by Fermi-Dirac statistics, while the ions are described by Maxwell-Boltzmann statistics. In this model, the screening from the bound electrons does not contribute to the IPD. We  note that multiple forms of the SP model have been described in the literature \cite{SP, JCP, Calisti, iglesias, crowley}. All SP calculations presented in this work use Eq.~\ref{eq:SP}. \\

The EK model is a generalized Saha equation as a function of the plasma's chemical potential. The model assumes two forms of the IPD, depending on the density of the ions and electrons combined. In the cases that we are investigating, the density is always above the ``critical density'', defined by the model as 
\begin{equation}
    n_c = \frac{3}{4\pi} \left( \frac{T}{z^2} \right)^3 \,.
\end{equation}
The form that the IPD takes above this critical density is given  by Eq.~\ref{Equation 18}. \\

\begin{figure}
    \centering
    \includegraphics[scale=0.17]{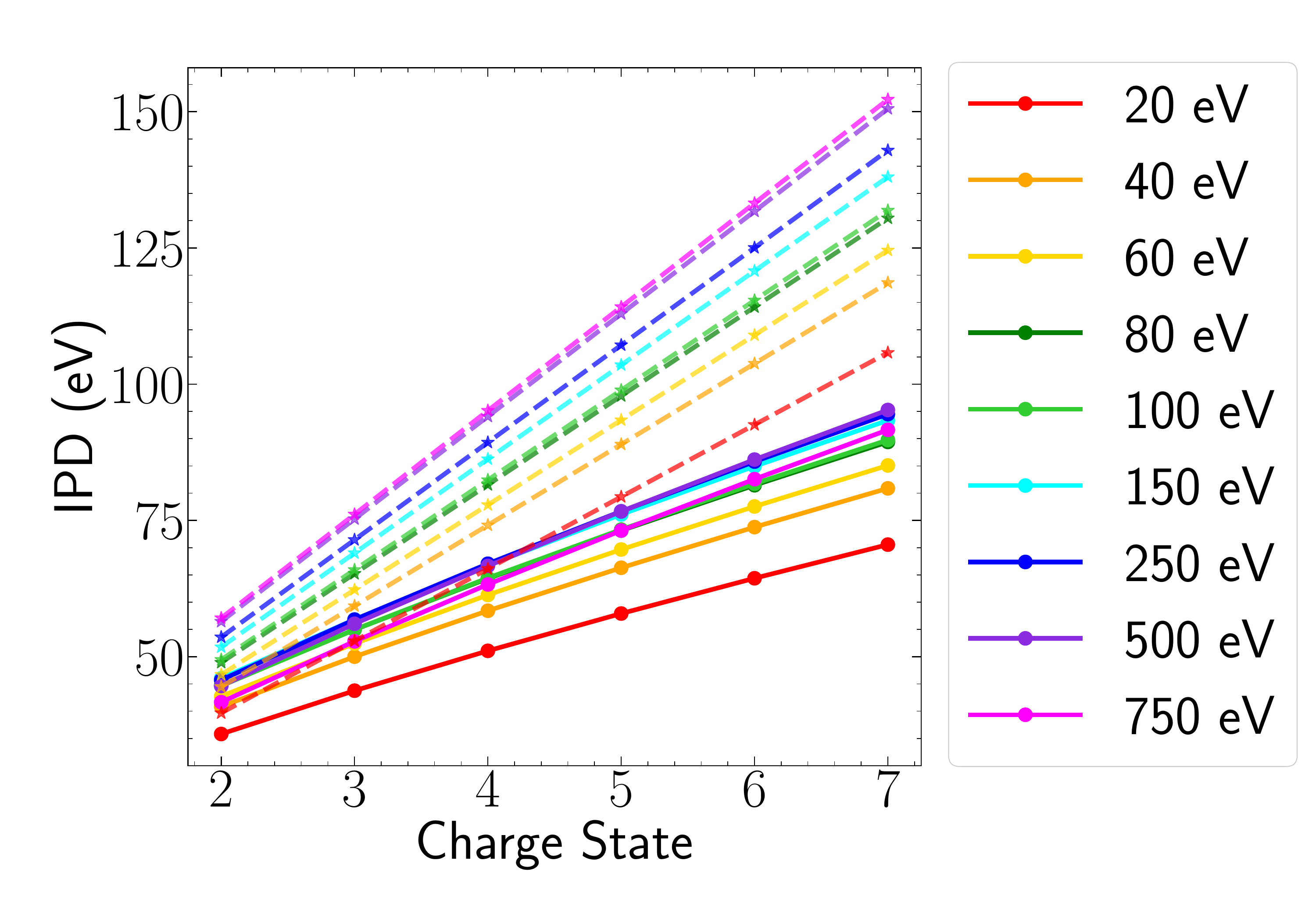}
    \caption[]{Comparison of the SP (solid lines) and the EK (dashed lines) models for temperatures ranging from 20 eV to 750 eV for solid density magnesium. For both models, the IPD increases as temperature increases. However, for the SP model, the IPD at a temperature of 750 eV is lower than that of 500 eV and 250 eV.}
    \label{fig:SP-EK-allT}
\end{figure}

Fig.~\ref{fig:SP-EK-allT} shows that there is a dependence on temperature for both the SP and EK models. Both models require an input of $z^*$, for which we have used the \texttt{Tartarus} average atom model \cite{starrett2019wide} to obtain. We can also see how the models compare to each other, e.g. the SP model  displays a shallower slope than the EK model. In the SP equation, the IPD  for a given charge state depends on two things: the temperature $T$ and the average ionization of the plasma $z^*$. For a constant temperature $T$, the IPD increases as $z^*$ increases. For a constant $z^*$, the IPD decreases as temperature increases. However, the value of $z^*$ increases as the temperature increases, though at high enough temperatures, the value of $z^*$ begins to plateau towards the maximum value (12 for magnesium). At these temperatures, $z^*$ can be considered constant compared to the change in temperature. This behavior causes the SP model IPD to decrease as $T$ increases for these higher temperatures. 

\begin{figure}
    \centering
    \includegraphics[scale=0.17]{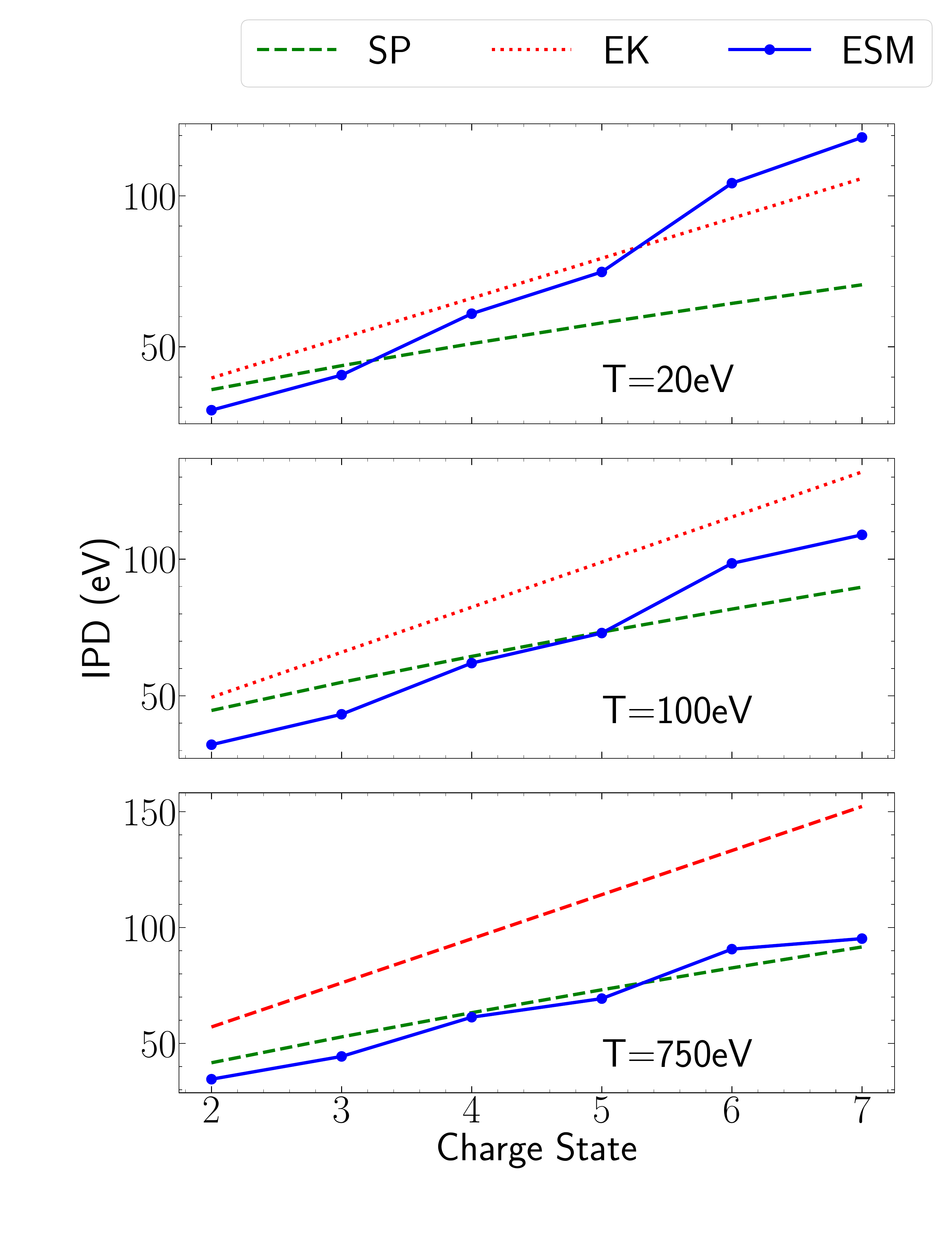}
    \caption{IPD calculated with our model compared with SP and EK models for magnesium at solid density and temperatures of 20 eV (top), 100 eV (middle), and 750 eV (bottom).}
    \label{fig:ipd-2temps}
\end{figure}

In Figure~\ref{fig:ipd-2temps} we compare the SP and EK predictions to the present excited states model at select temperatures. We find that at the lower temperatures, the IPD calculated with our model is slightly closer to the EK model, while at higher temperatures our model is closer to the SP model.

At temperatures consistent with experiments, the IPDs calculated with our model in Fig.~\ref{fig:ipd-2temps} have mixed agreement with the SP and EK models. This is consistent with the two-step Hartree-Fock results presented in Ref.~\cite{sks}, in which the calculated IPD falls between the SP and EK models, for aluminum in  that case. This behavior was also  reported in Ref.~\cite{crowley}. Additionally, Ref.~\cite{zan2021} shows  a model of the IPD of high density iron at $T\approx100$~eV that falls between the EK and SP models, which is consistent with our findings as well. 

Our results at $T=100$ eV for magnesium are consistent with the findings in Ref.~\cite{lin2019} for lower charge states, but disagree more at higher charge states, with our model predicting a lower IPD comparatively. Our magnesium results also disagree with the IPD predicted from the atomic-solid-plasma method \cite{rosmej2018}, which estimates an IPD greater than both the SP and EK models. 

\section{Conclusions}
We used our recently developed excited states model~\cite{starrett24excited} to model the experiments in Ref.~\cite{OC2012} and Ref.~\cite{oc16}. The model is variationally derived and uses an effective single-particle expression for the energy of a chosen set of excited states, defined by their single-particle occupation factors. The model was used to calculate K-edge and K-alpha energies, as well as the ionization potential depression for solid density magnesium, aluminum, and silicon at various temperatures.  

We found generally good agreement between the model and experiment for the K-edge and K-alpha.  Some differences were observed for higher charge states for the K-alpha energies. This behavior echoes earlier findings in Ref.~\cite{sks}. Some possible inadequacies in the present model that could cause this discrepancy are the absence of multiple scattering~\cite{starrett2020multiple} and the use of a local density approximation (LDA) expression for the exchange and correlation energy. 

Finally, we calculated the IPD over a large range of temperatures to compare to the SP~\cite{SP} and EK~\cite{EK} models, and also investigated the temperature dependence of the models themselves. As is shown in previous works, the EK model gives higher estimates of the IPD than the SP model, and depends more strongly on the charge state of the ion; both models also generally estimate an increase in IPD as the temperature increases.  We showed that the IPD calculated with our model at lower temperatures agrees more with the EK model, while our calculated IPD at  higher temperatures agrees better with the SP model, which is consistent with findings in Refs.~\cite{zeng2020} and \cite{JCP}. However, for the most part, our calculated IPD has mixed agreement with both models, rather than strictly agreeing with one model over the other, which is also found in other results in the literature \cite{sks, crowley, zan2021}. 

In summary, the new model can accurately predict the excitation energies in these dense plasmas. This level of agreement further validates our approach, while also revealing areas where additional research is needed.

\section*{Acknowledgements}
We thank Prof. S. Vinko for providing experimental data and useful discussions. 

This work was supported in part by the LDRD Program at Los Alamos National Laboratory under Project No. 20220172ER. This work was also supported in part by Advanced Simulation and Computing, Physics and Engineering Models, at Los Alamos National Laboratory. Los Alamos National Laboratory is operated by Triad National Security, LLC, for the National Nuclear Security Administration of the U.S. Department of Energy under Contract No.~89233218NCA000001.

\bibliography{test}

\end{document}